# Fourier plane optical microscopy and spectroscopy

Adarsh B. Vasista[1], Deepak K. Sharma[1] and G V Pavan Kumar[1,2,*]

[1] Department of Physics, Indian Institute of Science Education and Research, Pune-411008, India

[2] Centre for Energy Science, Indian Institute of Science Education and Research, Pune-411008, India

[*]**E-mail**: pavan@iiserpune.ac.in

## Abstract

Intensity, wavevector, phase, and polarization are the most important parameters of any light beam. Understanding the wavevector distribution has emerged as a very important problem in recent days, especially at nanoscale. It provides unique information about the light-matter interaction. Back focal plane or Fourier plane imaging and spectroscopy techniques help to measure wavevector distribution not only from single molecules and single nanostructures but also from metasurfaces and metamaterials. This review provides a birds-eye view on the technique of back focal imaging and spectroscopy, different methodologies used in developing the technique and applications including angular emission patterns of fluorescence and Raman signals from molecules, elastic scattering etc. We first discuss on the information one can obtain at the back focal plane of the objective lens according to both imaging and spectroscopy viewpoints and then discuss the possible configurations utilized to project back focal plane of the objective lens onto the imaging camera or to the spectroscope. We also discuss the possible sources of error in such measurements and possible ways to overcome it and then elucidate the possible applications.

## Table of Contents

# 1. Introduction

Fourier plane imaging , in other words , back focal plane (BFP) imaging and spectroscopy has emerged as an important tool, of late, in nanophotonics. It provides an alternative set of information which is not available in conventional real plane imaging. In a BFP image, one usually has information encoded in angular co-ordinates in contrast to the real plane image, where the resolution is in spatial co-ordinates. Measuring the angular emission pattern has been utilized in variety of applications including measuring radiation patterns from single quantum emitters (*1*-*5*), nanoantennas(*6*-*12*), understanding coupling mechanisms in cavities (*13*), in cathodoluminescence(*14*, *15*), secondary emission(*16*-*21*), scanning probe microscopy(*22*), soft-matter dynamics(*23*), non-linear scattering(*24*), full field and phase reconstruction(*25*, *26*), in understanding Kerker scattering(*27*), displacement sensing(*28*) and much more.

BFP imaging has been utilized in conoscopy (*29*-*31*) since early days. But the application of this technique to microscopy is a recent development, which needs attention. The advancements in micro/nanofabrication methodologies have made handling light at subwavelength scales possible. This calls for sophisticated microscopy techniques to understand and harness light at subwavelength scales. BFP imaging has emerged as an important tool along with scanning near field optical microscopy(*32*) , super-resolution microscopy(*33*), dark field microscopy(*34*), etc. Recently various approaches in BFP imaging has allowed to image different layers of samples using out of plane back focal plane imaging(*35*). Back focal plane microscopy has been applied to measure the effective refractive indices of the different modes in the system using Leakage radiation microscopy(*36*). In addition to this, BFP spectroscopy provides wavelength resolution to the angular information of the light emanated from the sample. This technique coupled to a microscope helps to measure energy-momentum dispersion relation of the outcoupled light from structures down to subwavelength limit (*13, 37-40*), angular dispersion in cavities with strong light-matter interaction(*41*) etc.

Here we explain, in detail, the concept of BFP imaging and spectroscopy, different methodologies and techniques involved in BFP measurement with a microscope, and its applications in nanophotonics, optical communication and nonlinear optics.

### 1.1. What is the back-focal plane of an objective lens?

In a typical microscope, the light emanating from the sample is collected using an objective lens and then routed to an imaging camera through external optics. Now, the light emerging from the sample plane can be expressed as weighted sum of different plane wave components, $u^i(x,y)$, emanating at different angles, $\theta^i$s. When this composite light passes through the objective lens, the different plane wave components get separated. The objective lens transforms the plane waves to paraboloidal waves which are focused at a point in the back focal plane. Any plane wave arriving at an angle $\theta_x$ , $\theta_y$ gets focussed onto ($\theta_x f$, $\theta_y f$ ), where $f$ is the focal length of the objective lens. The objective lens maps each direction ($\theta_x$,$\theta_y$) to a single point ($\theta_x f$, $\theta_y f$ ) onto the back focal plane and hence gives information about the angular distribution of the emanated light from the sample plane.

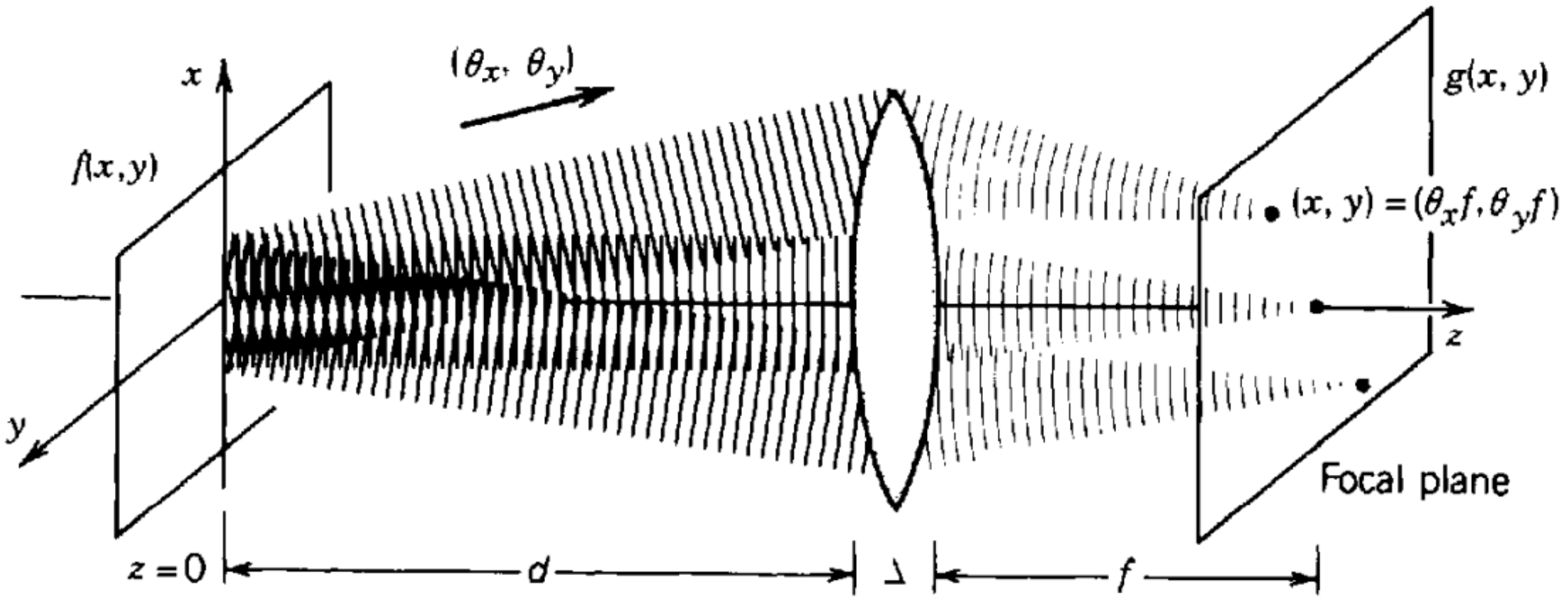

Fig1: Conceptual schematic of Fourier transform using a lens. Reproduced with permission from(*42*)

## 1.2. What information can we get using back focal plane of an objective lens?

### *1.1.1. Imaging view point*

The most important information which can be availed is the angular spectrum of the emanated light. In other words, far field emission pattern of the emission from a nano-object. As the angular distribution of light is intimately related to its momentum (h**k**), the BFP image can be viewed as inplane momentum ($k_x$,$k_y$) distribution of the outcoupled light. This enables to measure angle resolved scattering/emission from individual/array of nanostructures(*43*-*49*), perform single molecule orientation imaging and measure far field molecular emission patterns(*50*-*53*), designing optical/opto-mechanical sensors(*54*, *55*) etc.

### *1.1.2. Spectroscopy view point*

As the back focal plane has resolution in terms of inplane momentum vectors, one can select specific wavevectors and disperse them to obtain wavelength information. In other words, one can experimentally measure energy-momentum dispersion relation. BFP spectroscopy is an important tool to distinguish competing molecular processes at nanoscale(*13*), perform angle resolved absorption and emission spectroscopy(*12*, *56*-*58*) , measuring radiation patterns of Raman modes from a molecular system(*59*, *60*) etc.

# 2. Imaging the back focal plane.

As discussed above, the back focal plane of the objective lens contains information about the emission wavevectors from the sample plane. Imaging the BFP of the objective lens is tricky as physically the BFP of the objective lens lies near-inside the metallic case. Hence in a typical optical microscope, relay optics is utilized to project the BFP of the objective lens to the imaging camera.

## 2.1. How to image the BFP using an optical microscope?

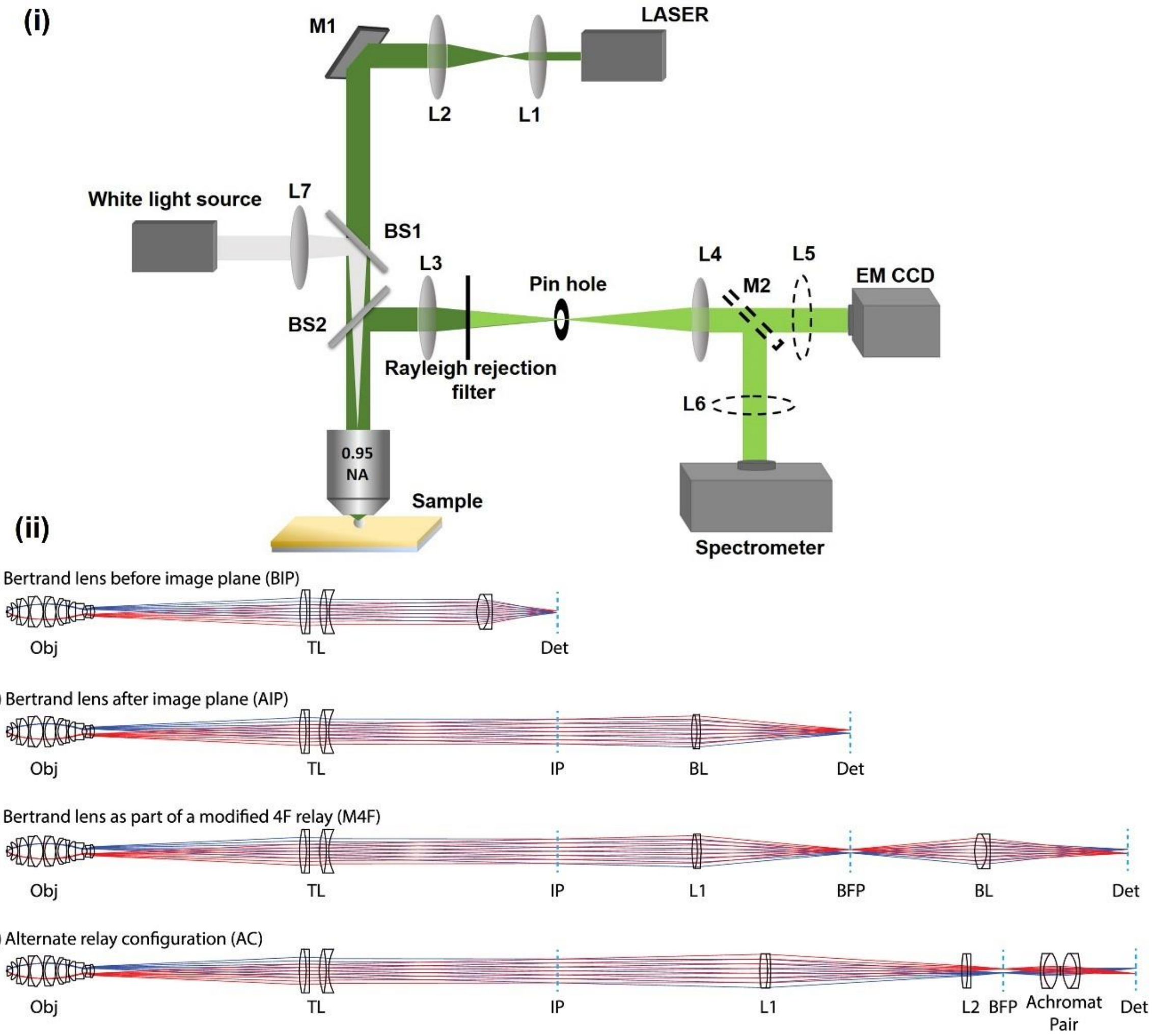

Fig. 2: (i) Schematic of imaging back focal plane using an optical microscope (ii) (a) – (d) Different relay configurations to image the back focal plane of the objective lens. *Legends:* NA: numerical aperture, BS: Beam splitter, IP: Image plane, TL: Tube lens, BL: Bertrand lens, Obj: Objective lens. Reproduced with permission from (*61*)(Optical Society of America).

Figure 2 (i) shows a typical setup used to perform back focal pane imaging and spectroscopy. This is a prototype setup and can be expanded and utilized with modifications. Here, the incoming laser beam is focused onto the sample plane using a high numerical aperture objective lens and the back scattered light is collected using the same objective lens (upright configuration). The collected light is then passed through lenses L3, L4 and L5 and projected to the CCD. The beam which is emanated from the infinity corrected objective lens is focused using lens L3, which creates a conjugate image plane. Lens L3 is placed at the distance of *f3*, the focal length of lens L3. Free optics like spatial filters can be placed at the conjugate image plane to filter out certain region of the sample. This configuration is very important as it can be extended for confocal microscopy, where a confocal pinhole is placed at the conjugate image plane in the place of spatial filter. Spectral filters like long pass / notch filters can be placed in the optical path to reject the incident laser, for secondary emission related measurements. This conjugate image plane now acts as the object plane for the lens L4, which projects the Fourier transform of the image plane onto its focal plane, where an imaging camera will be placed. Lens L4 is placed at a distance of *f4*, focal length of L4, from the pinhole. Thus the relay of lenses L3 and L4 projects the BFP of the objective lens onto the imaging camera outside the microscope. Combination of lenses L4 and L5 will create real plane image at the imaging camera. L5 is placed at a distance of *f5*, focal length of L5, from the CCD to focus the incoming beam to the CCD.

The configuration explained above is one among many such possible configurations to project the BFP of the objective lens onto the camera as seen in fig 2 (ii). Kurvits et.al.,(*61*) lists different possible configuration to project BFP onto the camera.

## 2.2. What are the possible sources of error and what are the limitations?

As different optical elements like lenses, mirrors and beam-splitters are placed in the optical path to image the BFP of the objective lens, there are different possibilities of generation of errors. Abberations (chromatic and spherical), vignetting, alignment issues, non-uniform intensity pattern of the BFP etc are possible errors in the imaging of BFP. We try to enumerate some possibilities. The list is not exhaustive.

### *2.2.1. Placement of the relay lenses*:

The tube lens essentially performs inverse Fourier transform and slight misplacement of the tube lens can actually make the phase information of the BFP lost (*42*). Since we usually measure intensity, where the phase information is integrated out, this will not be visualised. But the placement can have significance when BFP measurements are coupled with phase sensitive measurements. Also the choice of the focal length of the lenses play a crucial role in accurately projecting the BFP to the camera. Choice of very large (small) focal length of the lenses can create problem in relative placement of lens and magnification, thus affecting the formation of the conjugate BFP at the camera.

### *2.2.2. Chromatic and spherical aberrations:*

If the collection optical path is slightly off the optical axis of the lenses, then spherical aberrations can occur. This will affect the formation of BFP at the camera. To avoid this, proper care has to be taken during alignment to ensure that the collected light from the objective lens passes through the optical axis of the relay lenses. In the case of polychromatic light (white light imaging, imaging broadband secondary emission) chromatic aberrations can occur, which can be avoided by the use of achromatic lenses.

### *2.2.3. Choice of objective lens:*

Selection of objective lenses are also crucial in minimising aberrations in the imaging of the BFP. Kurvits et al.,(*61*) provides a detailed description for the choice of objective lenses for high numerical aperture imaging.

### *2.2.4. Placement of associated optics:*

Placement of bi-refringent optics like polarizers introduce in-plane displacement of the BFP at the imaging camera. Polarizers, in general, placed anywhere at the optical path can shift the position of BFP on the camera, which creates difficulty in measurements like BFP polarimetry. Hence polarizers need to be placed at the conjugate BFP. This conjugate BFP should be projected onto the camera using another set of relay lenses.

The information in the BFP imaging is limited by the numerical aperture of the collecting objective lens. Since the collection numerical aperture defines the collection angle, it is a very important parameter in designing the experiment. The resolution in the BFP is mainly defined by the magnification of the BFP at the imaging camera and the pixel size of the camera. The magnification of the BFP can be altered by changing the focal length ratio of Bertrand lens to the tube lens. For energy-momentum spectroscopy one needs relatively high intensity per pixel. This can be achieved by increasing the exposure time, input power etc or by decreasing the magnification of the BFP at the spectrometer slit. This can reduce the resolution further. So, there is a trade-off between the resolution one can obtain and the magnification of the BFP. The resolution of the E-k diagram is also limited by the grating of the spectrometer, slit width of the spectrometer etc.

# 3. Spectroscopy in back focal plane

In the previous sections, we discussed the importance and technique of performing BFP imaging. However when the collected light from the objective lens has specific dispersion relation, then it would be necessary to have wavelength resolution for the wavevector distribution. Spectroscopy in the BFP essentially provides the extra dimension needed for measuring dispersion relation. Here we project the BFP of the objective lens onto the spectrometer after filtering certain part of the BFP. Filtering can be achieved either by introducing a slit at the conjugate BFP(*37*) as shown in fig.3 or projecting the BFP onto the slit of the spectrometer and utilizing the slit as the filter(*13, 39*).

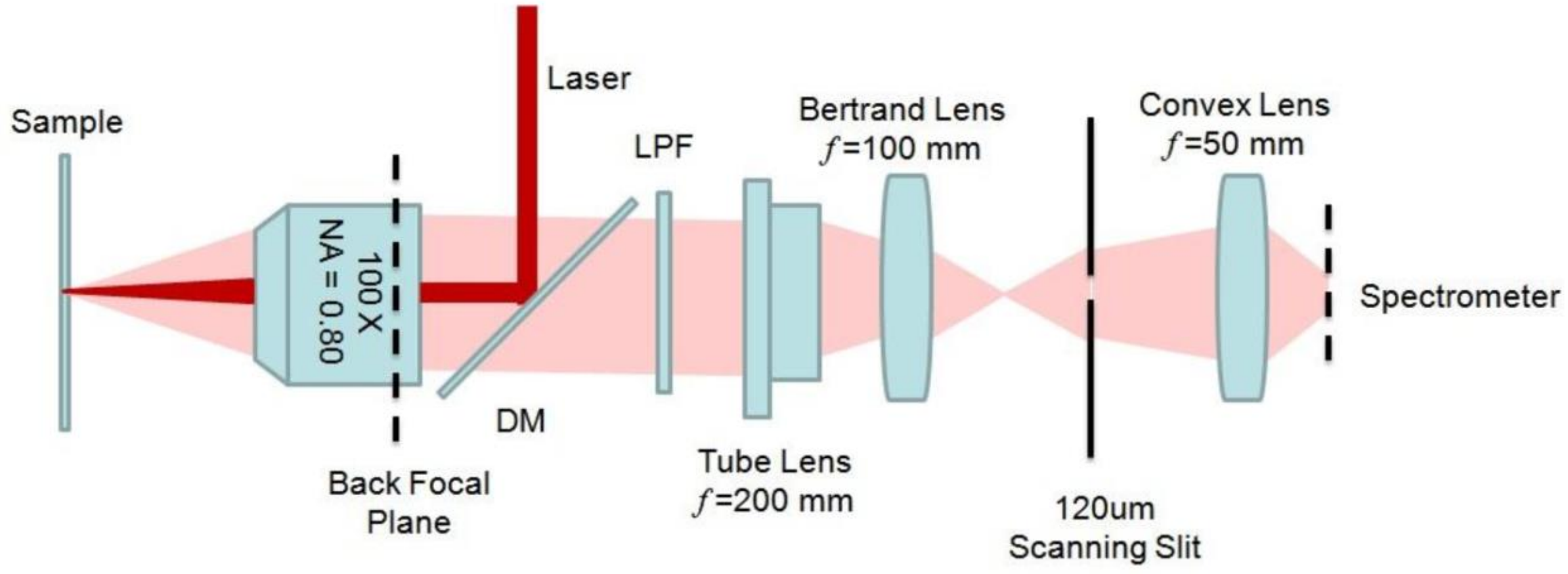

Fig.3: Schematic representation of back focal plane spectrometry. *Legends:* LPF: Low pass filter, DM: Dichroic mirror. Reproduced with permission from (*37*)

The information in a typical BFP image contains convolution of all possible wavelengths with angular resolution. If a process contains multiple modes of emission, like SERS (Surface enhanced Raman scattering), which can possibly occupy different wavevectors can only visualized by performing BFP spectroscopy (*18, 62*). Also such technique can be expanded to understand competing processes , Raman and fluorescence, from an ensemble of molecules(*13*) as these processes happen to emit at the same wavelength and cannot be separated in the real plane (in continuous wave limit). This provides an added advantage over real plane spectroscopy technique. The possible sources of error enumerated in the last section equally applies to the BFP spectroscopy also. In addition, the resolution of the spectrum now will also depend on the width of the slit used to filter the wavevectors in the BFP. Hence there will be a trade-off between the throughput and the resolution, which depends on the size of the slit, as in the case of real plane spectroscopy.

# 4. Applications

## 4.1. Single molecule fluorescence imaging

Back focal plane imaging is utilized extensively in molecular emission imaging and spectroscopy as intensity pattern in the back focal plane image is intricately related to the molecular orientation. Lieb et.al.,(*3*) has shown that back focal plane imaging can be utilized to probe single molecule orientations. By understanding the radiation patterns of molecular dipole single molecular orientations can be probed, as shown in figure 4.

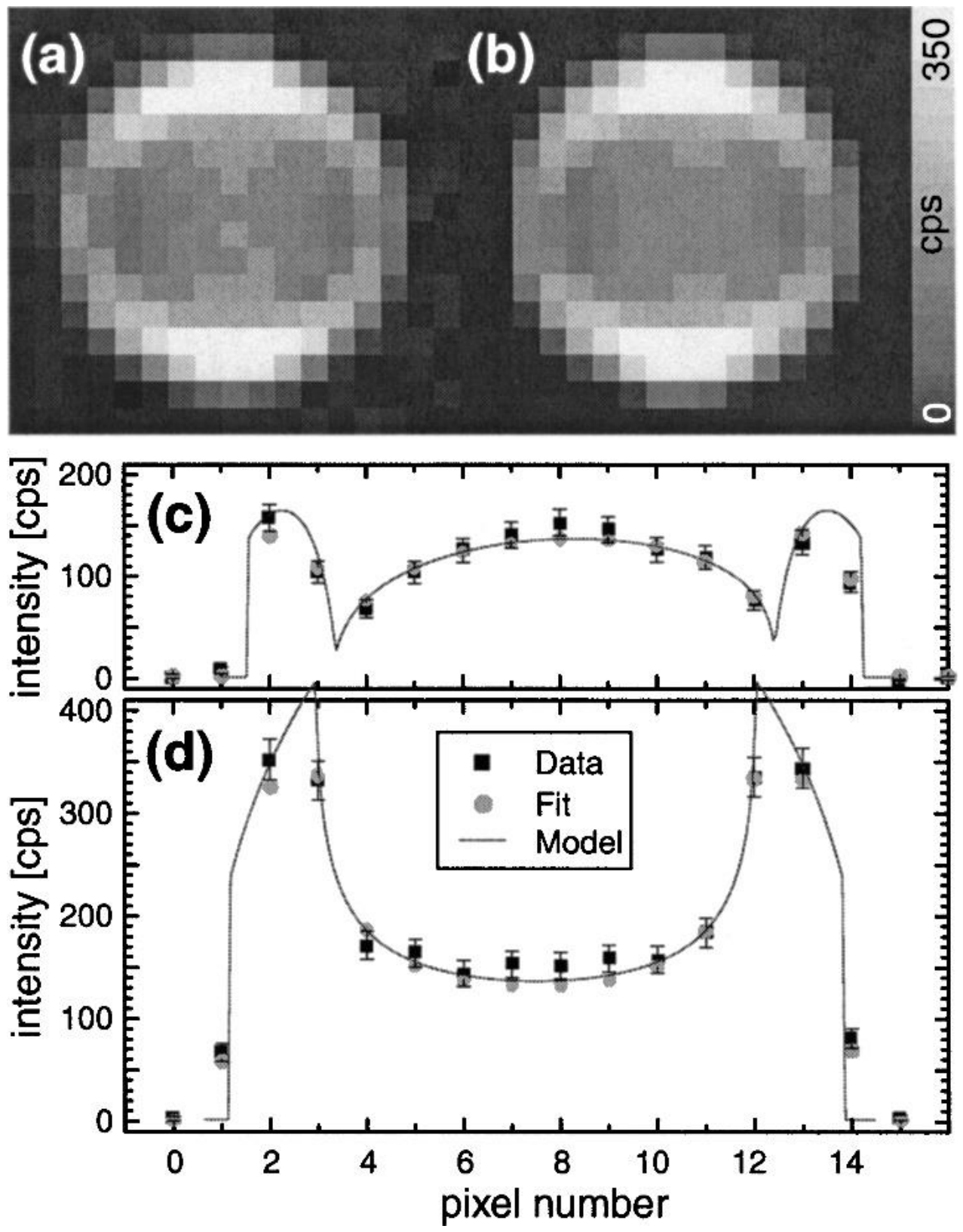

Fig.4: (a) Measured single molecule fluorescence emission pattern captured by imaging the back focal plane of the objective lens. (b) Fitted emission pattern of the molecular emission. (c) and (d) are intensity cross sections along the horzitonal and vertical line through the center of the pattern respectively. Reproduced with permission from (*3*)(Optical Society of America)

### 4.2. Raman scattering

Raman scattering provides unique information about the molecule. Angular distribution of the Raman scattering signals carry vital value, as one can determine molecular bonding orientation, interaction strength etc. Especially in anisotropic 2D materials like Graphene and Phosphorene, BFP images of different Raman modes is of crucial value. Budde et al.(*18*), have probed the angular radiation patterns from Graphene and demonstrated the important differences between the G and 2D bands of the material. BFP imaging coupled with output polarization resolution provides information about the polarizability of the material in addition to the angular radiation pattern of the modes of emission as shown in figure 5.

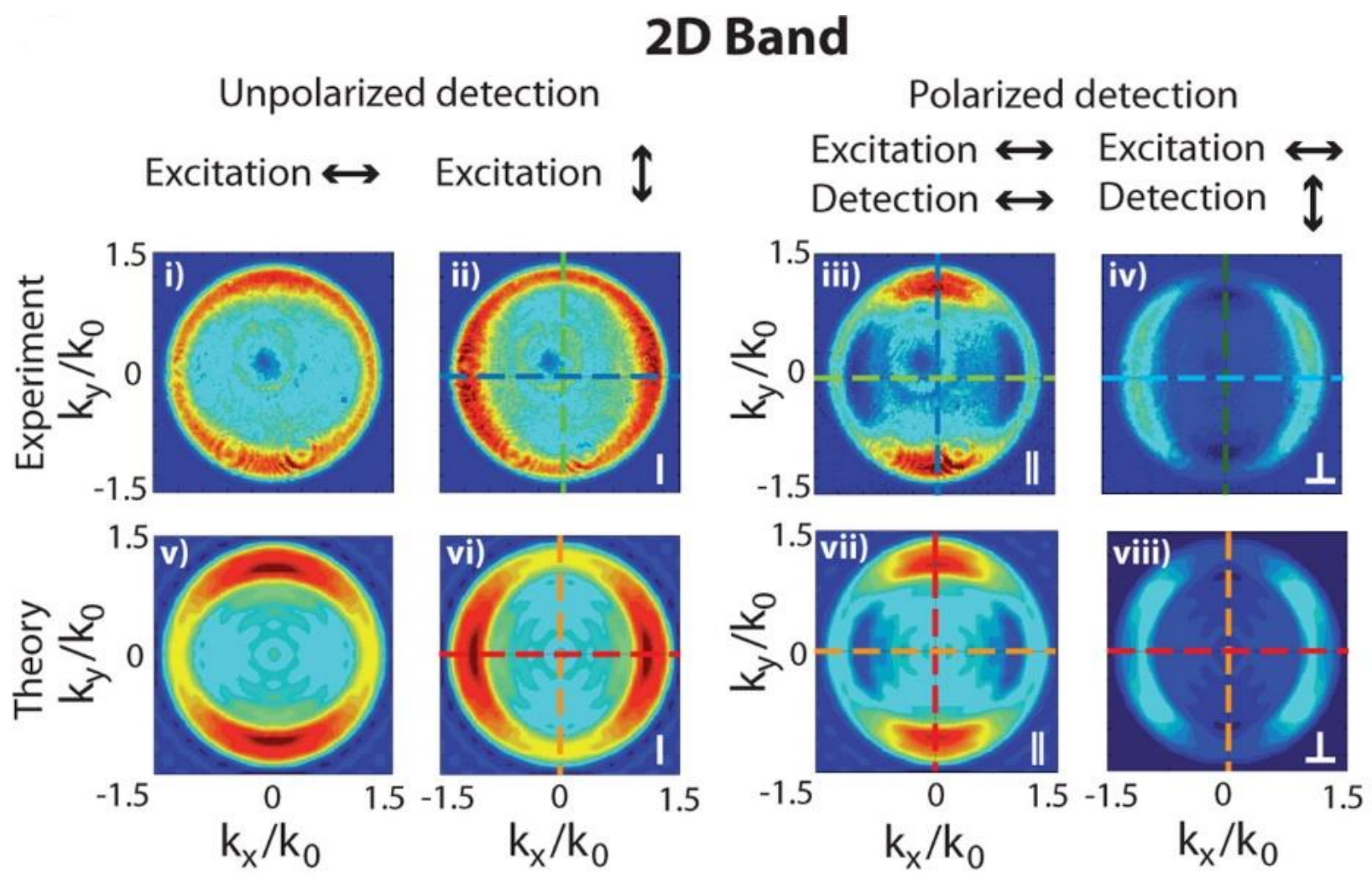

Fig.5: Back focal plane images captured by spectrally filtering 2D band of graphene, showing strongly polarized emission. Reproduced with permission from (*18*)

## 4.3. Plasmonic scattering/ Optical antenna

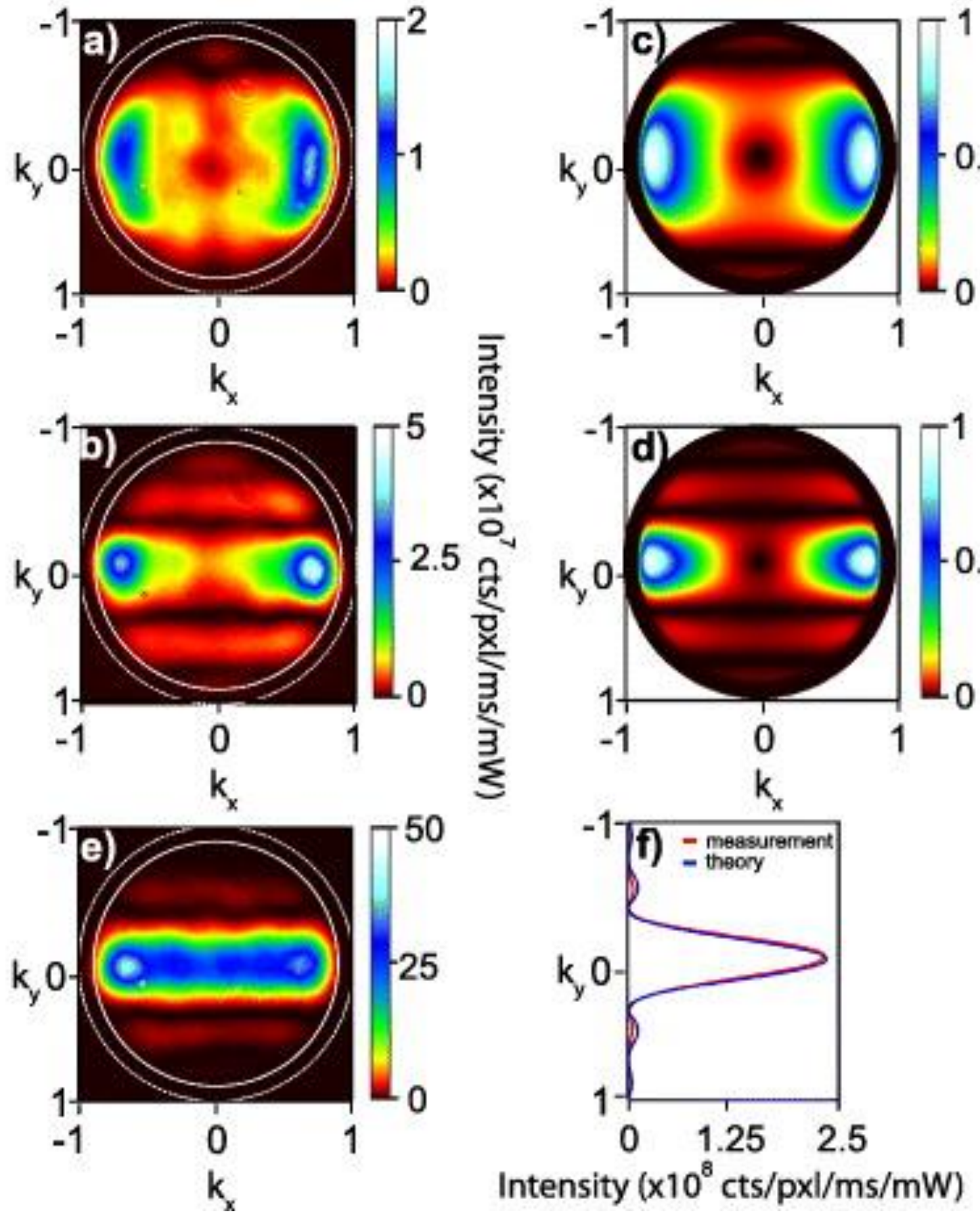

Fig.6: Back focal plane images captured from single plasmonic scatterers using evanescent field excitation. (a) and (b) are measured back focal plane images from 1 µm long and 2 µm long Au nanorods placed over glass substrate respectively . The width of the rod was 50 nm and excitation was p-polarized (c) and (d) are calculated BFP patterns. (e) measured BFP image for a 2 µm long, 50 nm wide Au nanorod on glass substrate with s-polarized excitation. Reproduced with permission from (*6*)

Farfield scattering of light from nano objects helps in understanding various optical phenomena at subwavelength scale which can further help in accommodating nanostructures in an integrated nano photonic device. This is essential as one of the important task in photonic circuits is to direct or guide light in particular direction and hence knowing the wave vector distribution is necessary. In the past, back focal plane microscopy has been utilized to understand elastic scattering from various nanostructures (*49, 63-69*). Sersic et.al (*6*) has shown how farfield scattering profile from a nanorod changes as a function of polarization of light and length of the nanorod, as shown in figure 6. It is clear from back focal plane images that farfield pattern changes drastically with the change in either geometry of nanorod or polarization of light.

On the other hand, Shegai et.al., (*8*) probed wave guiding properties of Ag nanowire and analyzed the outcoupled light using back focal plane microscopy. Back focal plane image in figure 7(c) reveals that outcoupled light from distal end of wire is unidirectional which is not evident by just looking at real plane propagation image of wire given in figure 7(b). Spread in the back focal plane image tells about the divergence of out coupled light which is an important information about a photonic device.

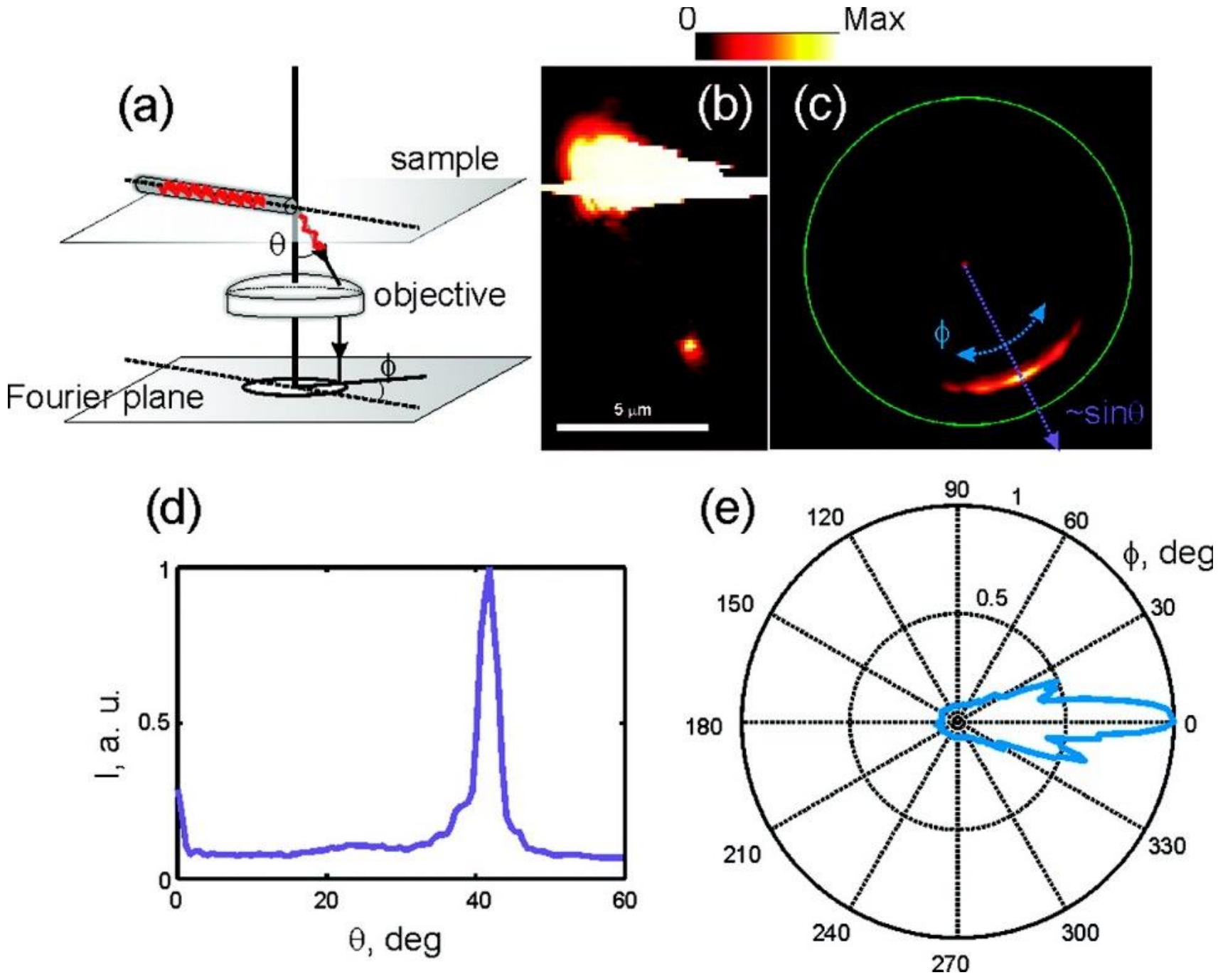

Fig.7: Unidirectional emission from single Ag nanowire. (a) Schematic of the emission configuration. (b) Real plane image showing the distal end of the wire. (c) Back focal plane image captured from distal end of an individual Ag nanowire. (d) and (e) are intensity profiles plotted along the radius of the BFP image and circumference at the maximum emission angle. Reproduced with permission from (*8*)

## 4.4. Secondary emission via optical antenna

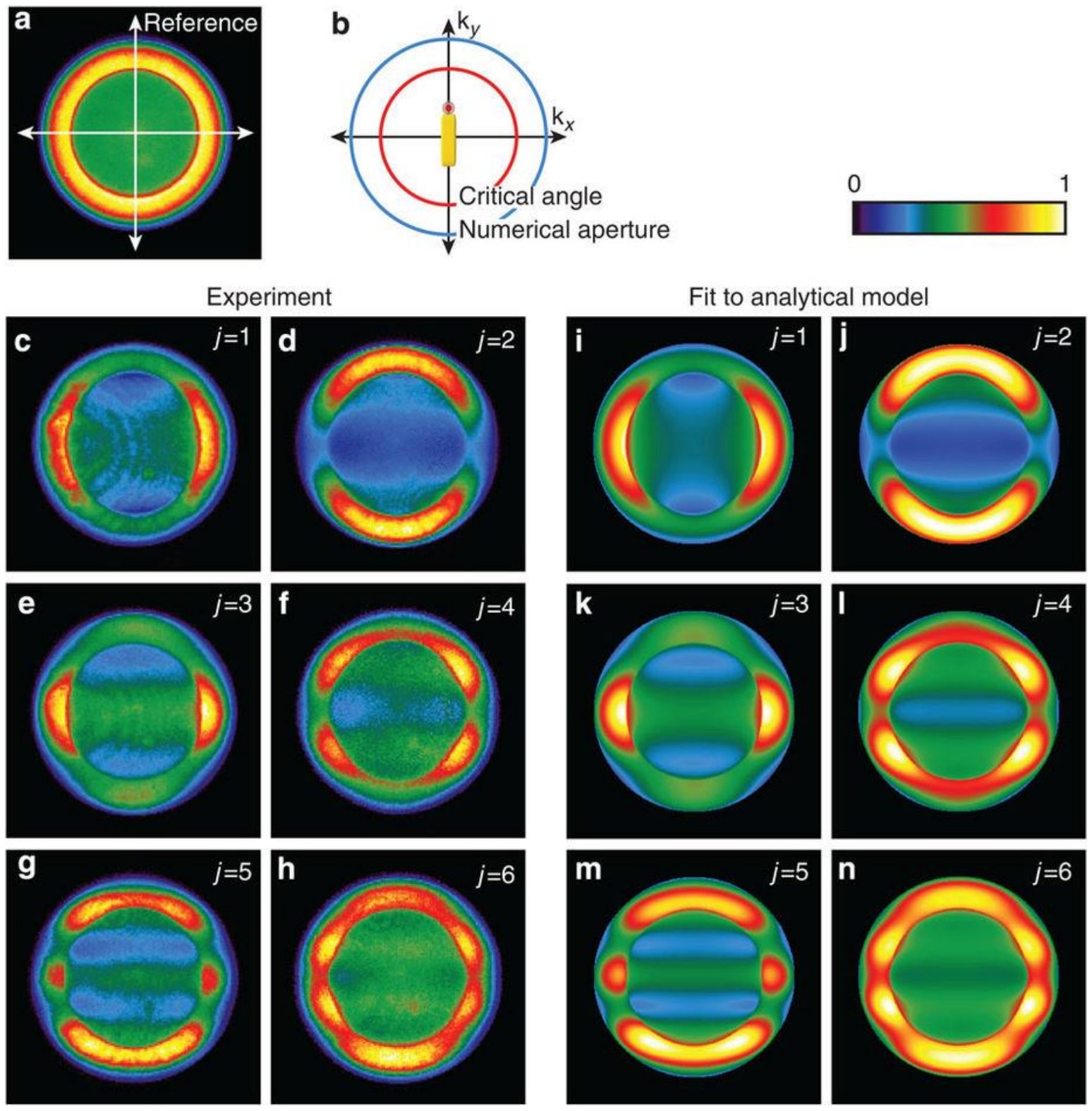

Fig.8 : Back focal plane fluorescence images captured from single quantum dot coupled to plasmonic nanorod. (**a**) Reference structure (**b**) Schematic of an angular pattern highlighting the critical angle of the glass–air interface and the numerical aperture of the objective (maximum collection angle). (**c**–**h**) Experimental patterns of the first six resonant antenna modes. (**i**–**n**) The patterns are reproduced with the analytical resonator model, including a rotationally symmetric background. Reproduced with permission from (*1*)

We discussed about probing molecular emission process using back focal plane imaging. This technique has been extended quite extensively to study molecular emission coupled to nanostructures. BFP imaging technique is of critical importance in probing antenna effects in plasmon coupled molecular emission. Recent reports (*70-73*) show the importance of BFP imaging in studying plasmon coupled molecular fluorescence. BFP imaging not only provides information about the emission properties of the molecule but also that of the plasmonic nanostructure, as studied by (*1*).

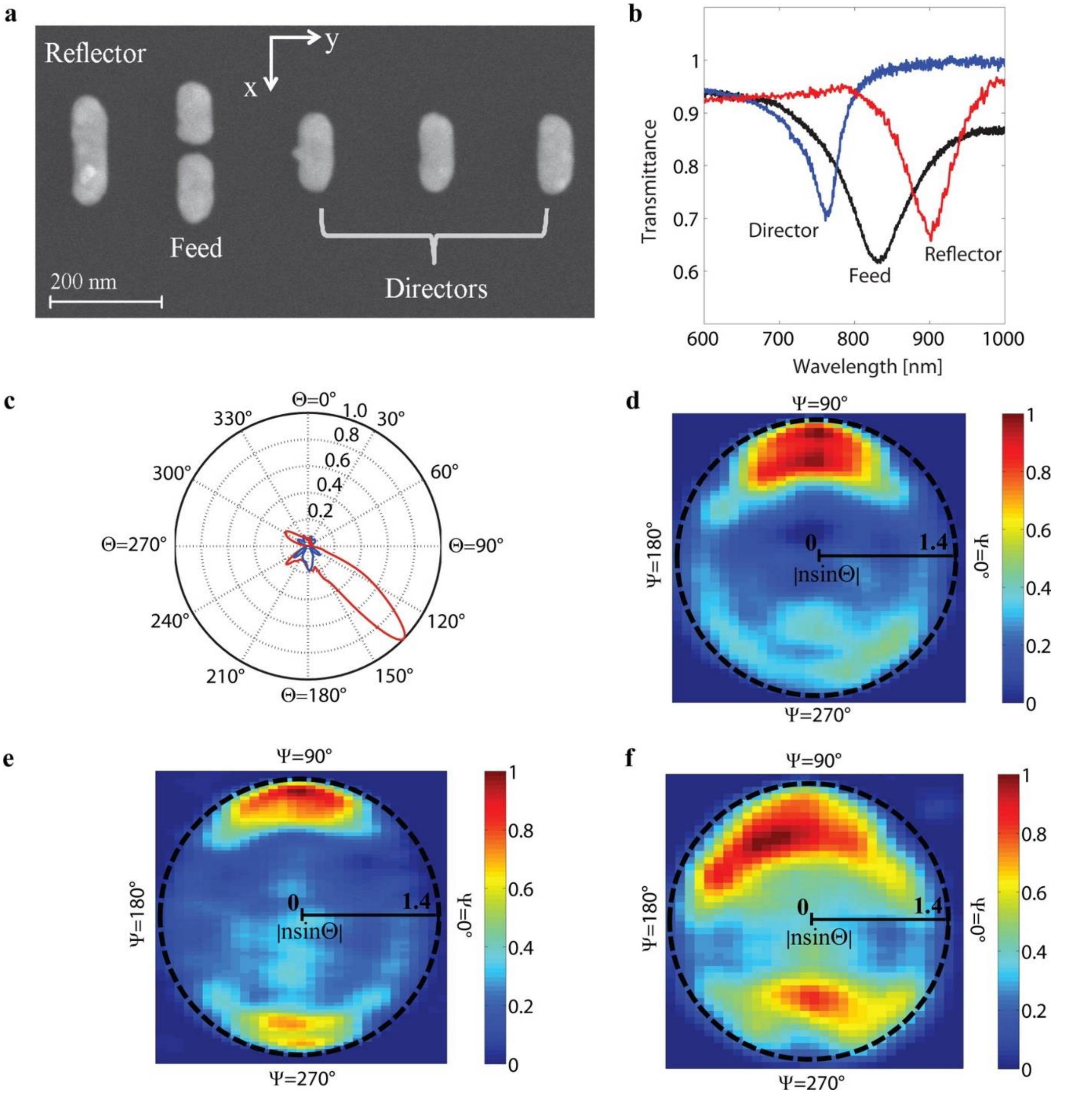

Fig.9: SERS emission pattern from Yagi Uda antenna. (a) SEM image of the Yagi Uda antenna. (b) Measured transmittance of arrays of directors, feed antennas, and reflectors. (c) FEM simulation of emission pattern resulting from an electric dipole (free space wavelength λ = 857 nm) being placed in center of gap in feed element of YU antenna; red: emission pattern in *yz*-plane and blue: emission pattern in *xz*-plane. (d) Emission pattern of thiophenol 1074 $cm^{-1}$ Raman line retrieved from E-k measurements. Measured emission patterns of (e) 415 and (f) 1586 $cm^{-1}$ Raman lines. Reproduced with permission from (*37*)

In the similar manner molecular vibrational transitions, especially Raman, gets affected when the molecule is coupled to a nanostructure. Wavevector distribution of SERS, in such cases, can alter drastically. BFP imaging and spectroscopy has been utilized to understand SERS with particular emphasis to wavevector distribution, as shown in figure 9(*37*). This has been further extrapolated to study coupling strengths and molecular orientations, which has applications in quantum electrodynamics.

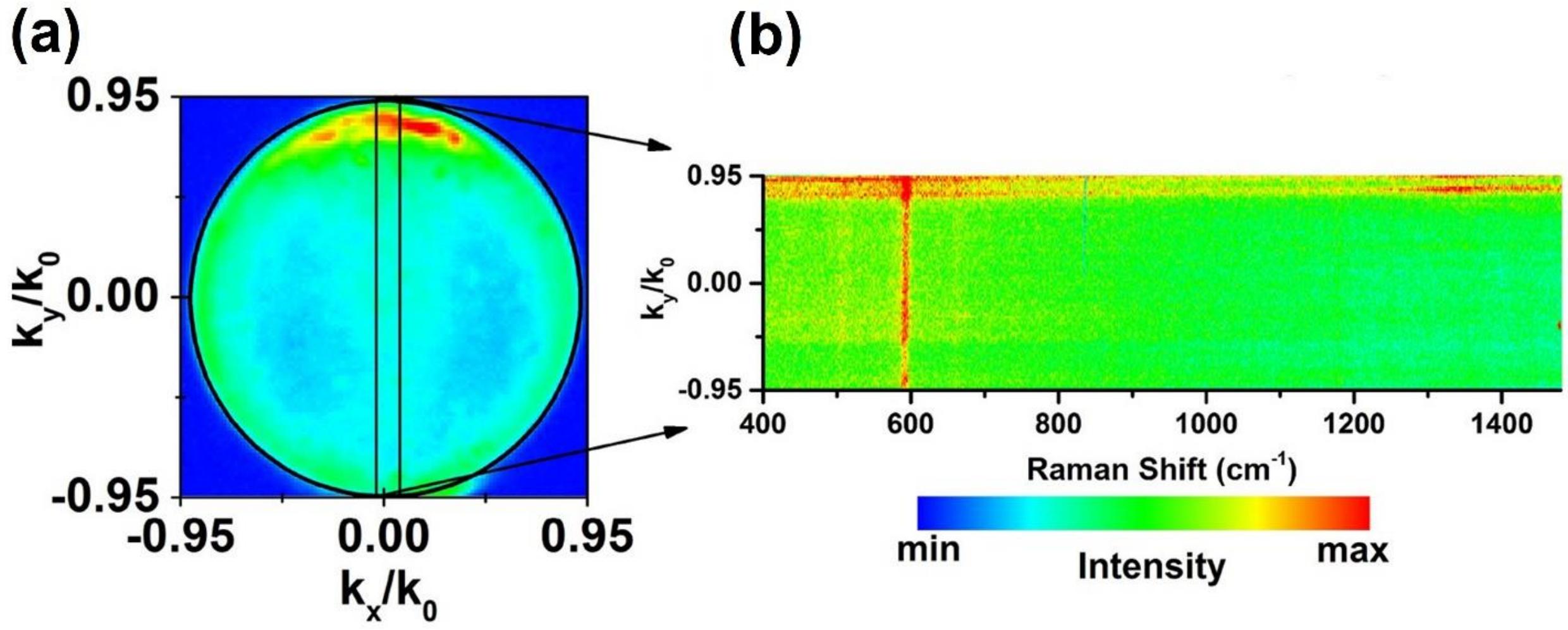

Fig.10: Imaging remote excitation SERS radiation pattern by filtering narrow range of wavevectors from the BFP. (a) Back focal plane image captured from the distal end of a film coupled plasmonic nanowire cavity, showing unidirectional emission pattern. (b) Energy-momentum spectrum captured by filtering a narrow range of the back focal plane image near $k_x/k_0=0$. Reproduced with permission from (*13*)

Another important aspect of BFP imaging is to understand the coupling of molecules to nanocavities. Nanocavities have specific wavevector signatures which can alter the emission wavevectors of the molecules coupled to such cavities. This inturn provides a unique way to measure the coupling strength and efficiency of molecules. Certain nanocavities, eg. Elongated cavities like film coupled nanowire cavity, have very complex polarization signatures. This makes polarization sensitive emission process like SERS couple differently to the system when compared to molecular fluorescence, as shown by Vasista et.al.(*13*).

### 4.5. Nonlinear scattering

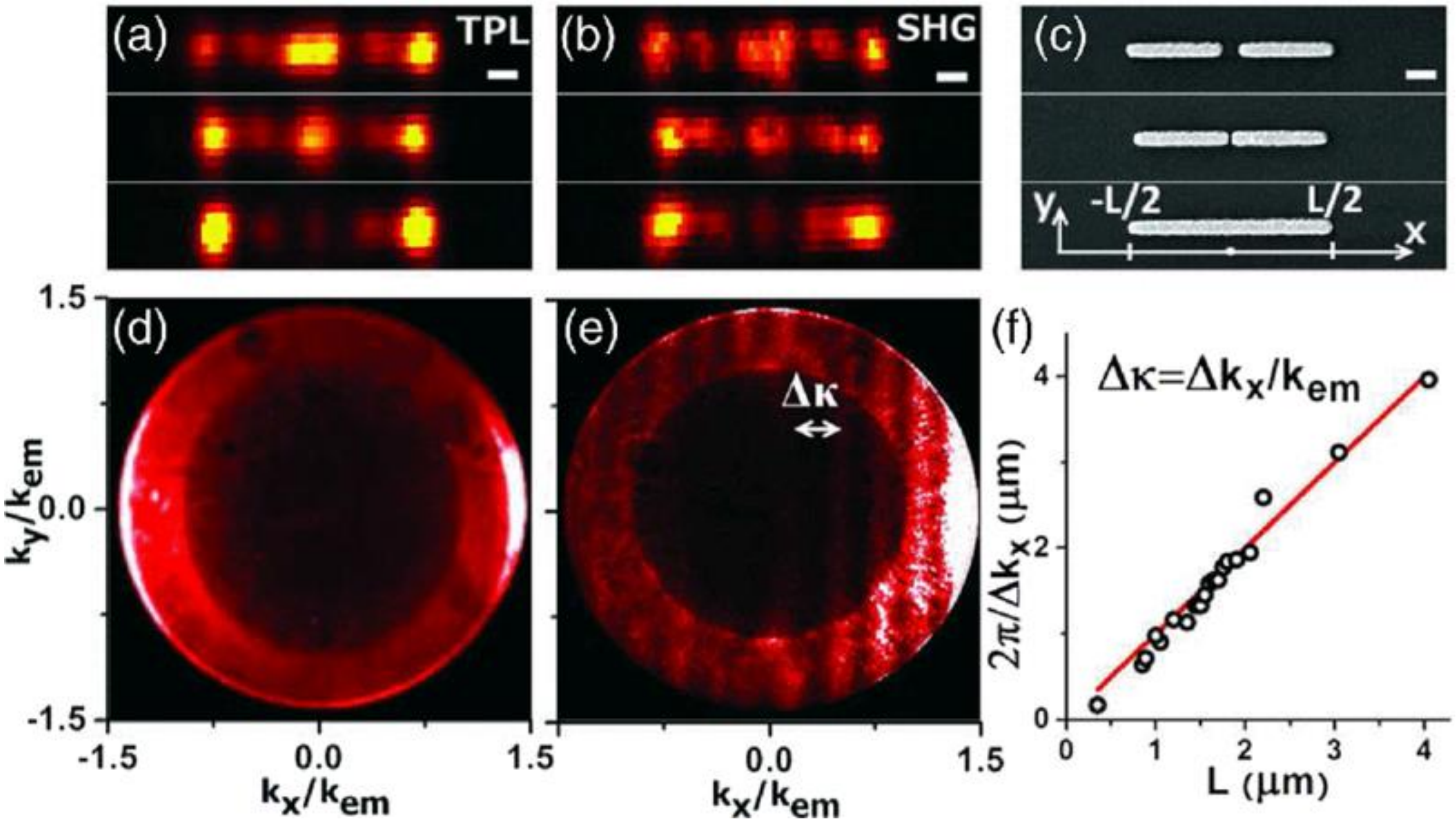

Fig.11: Back focal plane images of Nonlinear signals from Au nanoantenna. a)–(c) TPL and SHG confocal maps and SEM micrographs of three gold rod antennas. (d) and (e) are angular distributions of TPL and SHG signal from a single nanowire [bottom antenna in (a)–(c)]. The laser is focused at the left extremity. (f) Dependence of $2\pi/\Delta k_x$ on L (black ∘) and $2\pi/\Delta k_x$=L (line). The axes are normalized in units of $k_{em}$=nm$2\pi/\lambda_{em}$, where $\lambda_{em}$ is the emission wavelength of the nonlinear process in vacuum and nm=1.55 is the refractive index of glass and oil. The interfringe distance Δκ is denoted with the white double-headed arrow. Reproduced with permission from (*74*)

Depending on the generation nonlinear signals can be incoherent like two photon luminescence (TPL) or coherent like second harmonic generation (SHG), third harmonic generation etc. Both coherent and incoherent nonlinear signals have unique spectral properties and hence are distinguishable spectrally. But just observing the real plane images (shown in Fig. 11 (a) and (b)) one cannot tell about the far field emission patterns. Back focal plane images shown in figure 11 (d) for TPL and figure 11 (e) for SHG shows that two signals have different far field patterns. Formation of fringes in back focal plane in figure 11 (e) again confirms the coherent nature of the SHG(*74*).

# 5. Conclusions

In summary, we discussed the technique of back focal plane imaging and spectroscopy in detail. We have also pondered upon different applications of the technique with specific emphasis on recent developments. Back focal plane imaging and spectroscopy forms a crucial factor in understanding emission processes like Raman spectroscopy(*18*), molecular fluorescence (*75, 76*), bio-imaging(*77*), non linear scattering(*78*), to understand strong coupling(*79*) etc along with elastic scattering by nano/micro

structures(*80*). The knowledge of Fourier plane microscopy has recently been extended to study structured light in nanoscale, specifically by Peter Banzer's group(*25, 81*). This knowledge can be further extrapolated to quantum optics and quantum electrodynamics to understand coupling mechanisms between atom and cavity etc.

# Suggested further reading